\date{100408}
\documentclass[showpacs,twocolumn]{revtex4}
\usepackage{graphicx,psfrag,amsmath,amssymb,amsfonts,latexsym,color,dcolumn,
graphpap}

\definecolor{red}{rgb}{1,0,0}
\definecolor{blue}{rgb}{0,0,1}
\definecolor{skyblue}{rgb}{0,0,.5}
\definecolor{green}{rgb}{0,1,0}
\definecolor{orange}{cmyk}{0,.4,1,0}

\begin{document}

\title{Computing the Casimir  energy using the point-matching method}

\author{F.C. Lombardo$^1$ \footnote{lombardo@df.uba.ar}}
\author{F.D.  Mazzitelli$^1$\footnote{fmazzi@df.uba.ar}}
\author{M. V\'azquez$^2$ \footnote{mariano.vazquez@bsc.es}}
\author{P.I. Villar$^{1,2}$ \footnote{paula@df.uba.ar}}

 \affiliation{$^1$ Departamento de F\'\i sica {\it Juan Jos\'e
 Giambiagi}, FCEyN UBA, Facultad de Ciencias Exactas y Naturales,
 Ciudad Universitaria, Pabell\' on I, 1428 Buenos Aires, Argentina \\
$^2$ Computer Applications on Science and Engineering Department,
Barcelona Supercomputing Center (BSC),
29, Jordi Girona
08034 Barcelona,
Spain}

\date{today}
\begin{abstract}
We use a  point-matching approach
 to numerically compute the Casimir
interaction energy for a two perfect-conductor waveguide
of arbitrary section.
We present the method and describe the procedure
used to obtain the numerical results.
At first, our technique is tested for geometries
with known solutions, such as concentric and eccentric cylinders. Then, we
apply the point-matching technique to compute the Casimir interaction energy
for new geometries such as concentric corrugated cylinders and cylinders
inside conductors with focal lines.
\end{abstract}


\maketitle

\section{Introduction}

One of the most clear macroscopic manifestations of quantum mechanics
is the Casimir force \cite{Casimir}:
a tiny force, relevant at short
distances, which appears on uncharged bodies due to changes in
the zero-point energy associated to quantum vacuum fluctuations.
Quantum effects like Casimir forces have become increasingly
important as electronic and mechanical systems on the nanometer
scale become more prevalent.
In the last 10 years, Casimir-force measurements have been made
in a wide variety of experiments. For example, plate-plate
\cite{Bressi}, sphere-plate \cite{Krause} and crossed cylinders
\cite{Ederth} configurations. There are also proposals 
for measuring the Casimir force in the cylinder-plane 
configuration \cite{Brown}. A recent experiment \cite{Ardito}
implies that Casimir interactions are relevant in the fabrication
of commercially MEMs devices.
Techniques for predicting Casimir forces in general
geometries are clearly needed if theory is to keep pace with
the flourishing of experimental Casimir data and the design challenges
of future MEMs devices.

In this framework, accurate modeling of Casimir forces in general
geometries and for arbitrary electromagnetic properties of the
uncharged bodies  becomes very important.
Shape and geometry can strongly influence the Casimir interactions.
Several theoretical techniques have been developed in order to understand the geometric dependence of the Casimir force. These include the use of the argument theorem to perform explicitly the sum over modes \cite{saharian,Mazzitelli2003,PRARC,NJP},
semiclassical and optical  approximations \cite{semiclass}, methods based on functional integrals
\cite{funct} and scattering theory \cite{scatt}. Many of these approaches have a common root
in the multiple scattering theory developed in the seventies \cite{Balian} (see also \cite{milton wagner} for an updated review and applications to semitransparent bodies), and
the  evolution in the computational power allowed  a
 precise numerical evaluation that involves, in general, the computation of determinants of
 infinite matrices.

There are also full numerical approaches, as the worldline
numerics \cite{gies},  that has been applied to scalar fields satisfying Dirichlet boundary
conditions, or finite difference methods that evaluate the Casimir energy from the
two-point function of the electromagnetic field \cite{Rodriguez}.
As a consequence of this theoretical activity, we
now have exact results for a variety of geometries that involve perfectly conducting shells:
cylinder and sphere in front of  a plane \cite{Emig,emig,paulo},
eccentric cylinders \cite{PRARC,NJP,PRD}, two-spheres \cite{emig,Emigspher}, surfaces with periodic corrugations \cite{emig3}, Casimir pistons \cite{pistons},   multibody interactions involving plates and cylinders \cite{Rahi}
or squares \cite{Rodriguez}, objects of spheroidal or nearly-spheroidal shape \cite{spheroid}, etc. 
Some of these methods
also apply to the case of imperfect mirrors.

The Casimir force between two conductor bodies with simple shapes is in general attractive
and monotonically decreasing with the separation among conductors.
Thus, one might ask  whether complex geometries might give
rise to unexpected  phenomena, such as non-monotonic
forces. For more
complicated geometries, however, full numerical calculations become
extremely difficult and therefore it is worth to analyze alternative approaches.

In this paper, we explore a new numerical approach to compute
the Casimir interaction energy in a two conductor waveguide
for arbitrary cross section.
The idea is to combine a well known method for computing eigenvalues
of the Helmholtz equation, the point-matching technique \cite{pmm}, with the argument
theorem to compute explicitly the sum over eigenfrequencies.
Our technique is first tested for geometries
with known solutions and then is applied to new geometries.
Thus, in Section \ref{method} we will
describe the new approach.  In Section \ref{ecc} we validate our method
against numerical results \cite{NJP, PRD} obtained by  us
in previous works.
In the rest of the paper, we use our numerical approach to
compute the Casimir interaction energy for more general
cross sections of the waveguide, as corrugated cylinders in
Section \ref{rack}. In Section \ref{elipses} and \ref{parabola},
we explore the behaviour of the Casimir interaction energy between
outer conductors that have focal lines and a cylindrical inner conductor. Concretely,
Section \ref{elipses} shows the Casimir interaction energy
when a cylinder with circular section is placed inside an outer
cylinder with elliptical section. In Section \ref{parabola}, the
outer conductor is a waveguide with parabolic section.
Finally, in Section \ref{conc} we expose our final remarks.

\section{Point-Matching Numerical Approach}
\label{method}
A two-conductor waveguide presents an interesting setup for the application
of the point-matching technique. This technique
has been widely used to solve
eigenvalue problems in many areas of engineering science
\cite{pmm}.
The boundary conditions are imposed at a finite number
of points around the periphery of both conductors. Under this assumption,
and for a solution proportional to $e^{-i\omega t}$,
the partial differential equation of the problem can be
reduced to a system of linear algebraic equations. The determinant
associated to this system vanishes for some values of $\omega$,
the eigenfrequencies of the system, that can be determined in this way by searching
numerically the zeros.
In order to obtain the Casimir energy, instead of computing each eigenvalue, it is more
convenient to use the argument theorem to perform the sum over all eigenvalues.
\begin{figure}[ht]
\centering
\includegraphics[width=6.cm]{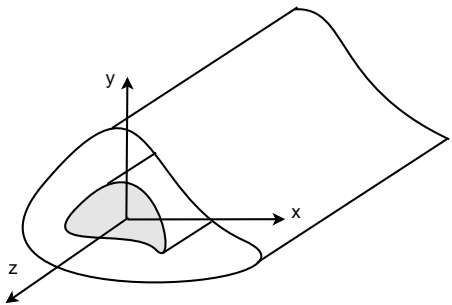}
\caption{A two-conductor waveguide in which one
conductor encloses the other and each has arbitrary
cross section.}
\label{fig1}
\end{figure}

Let us consider a general geometry with translational invariance
along the $z-$axis (as for example very long and parallel conducting shells
of arbitrary sections).  We will bear in mind the situation in which
one conductor encloses the other, as shown in Fig. \ref{fig1},
although the method could be applied to more general cases. The
Casimir interaction energy for a system like this, composed of two conducting shells,
can be written as
\begin{equation}
E_{12}= \frac{1}{2} \sum_p(w_p-\tilde w_p) ,
\label{ecasmodes}
\end{equation}
where $w_p$ are the eigenfrequencies of the electromagnetic field
satisfying perfect conductor boundary conditions on the surfaces
of the conductors, and $\tilde w_p$ are those corresponding to 
a situation in which the external conducting shell  
is very large.
Throughout this
paper we use units $\hbar=c=1$. The subindex
$p$ denotes the set of quantum numbers associated to each
eigenfrequency. Introducing a cutoff for high frequency modes
$E_{12}(\sigma)={1\over 2}\sum_p(e^{-\sigma w_p} w_p-e^{-\sigma
\tilde w_p} \tilde w_p)$,
the Casimir interaction energy $E_{12}$ is the limit of $E_{12}(\sigma)$ as
$\sigma\rightarrow 0$. For simplicity we choose an exponential
cutoff, although the explicit form is not relevant.

The transverse electric (TE) and
transverse magnetic (TM) modes can be described in terms of two
scalar fields,  with adequate boundary conditions.
In cylindrical
coordinates, the modes of each scalar field will be of the form
$h_{n, k_z}(t,r,\theta,z)=e^{(-iw_{n, k_z}t+ik_z
z)}R_n(r,\theta)$, where the eigenfrequencies are $w_{n,
k_z}=\sqrt{k_z^2+\lambda^2_n}$, and $\lambda_n$ are the
eigenvalues of the two dimensional Laplacian
\begin{equation}
\left(\frac{\partial^2}{\partial
r^2}+\frac{1}{r}\frac{\partial}{\partial r}+
\frac{1}{r^2}\frac{\partial^2}{\partial\theta^2}+\lambda_n^2\right
) R_n(r,\theta)=0.\label{helmho}
\end{equation}
The set of quantum numbers $p$ is given by
$(n, k_z)$. For very long cylinders of length $L$ we can replace
the sum over $k_z$ by an integral. The result is
\begin{eqnarray}
E_{12} (\sigma) &=& {L \over 2}\int_{-\infty}^{\infty}{dk_z\over
2\pi} \sum_{n}\left (\sqrt{k_z^2+\lambda_{n}^2}e^{-\sigma
\sqrt{k_z^2+\lambda_{n}^2}} \right.\nonumber \\
 &-& \left. \sqrt{k_z^2+\tilde \lambda_{n}^2}
e^{-\sigma \sqrt{k_z^2+\tilde\lambda_{n}^2}} \right) \; .
\label{exs}
\end{eqnarray}
From the argument theorem it follows that
\begin{equation}
{1\over 2\pi i} \int_{C} \,d\lambda \;  \lambda \; e^{-\sigma \lambda} {d\over d\lambda}
\ln f(\lambda)=\sum_i \lambda_i \; e^{-\sigma \lambda_i} \; ,
\end{equation}
where $f(\lambda)$ is an analytic function in the complex $\lambda$ plane within the closed contour
${C}$, with simple zeros at $\lambda_1, \lambda_2, \dots$ within ${C}$.
We use this result to replace the sum over $n$ in Eq.(\ref{exs}) by a contour integral
\begin{eqnarray}
E_{12}(\sigma) &=& {L\over 4\pi i}\int_{-\infty}^{\infty} {dk_z\over 2\pi}
\int_{C} d\lambda \sqrt{k_z^2+\lambda^2}
e^{-\sigma \sqrt{k_z^2+\lambda^2}} \nonumber \\
&\times & {d\over d\lambda} \ln Q(\lambda) \; .
\end{eqnarray}
Here the function $Q(\lambda)$ is the ratio $Q(\lambda)=F(\lambda)/\tilde F(\lambda)$ such that $F(\lambda)$ vanishes at $\lambda_n$ 
and $\tilde F(\lambda)$ vanishes at $\tilde\lambda_n$ for all $n$.

To proceed we must choose an adequate contour for the integration in the
complex plane (see \cite{NJP} for details). We find
\begin{equation}
E_{12}= -{L \over 2\pi}\int_{-\infty}^{\infty} {dk_z\over 2\pi} ~{\rm Im}
\left\{\int_0^{\infty} dy \sqrt{k_z^2-y^2} {d\over dy}\ln Q(iy)
\right\}. \label{xx}
\end{equation}
As we will see, $Q(iy)$ is a real function hence,  the integral
over $y$ in Eq. (\ref{xx}) is restricted to $y > k_z$. After some straightforward
steps one can re-write this equation as
\begin{equation}
E_{12}={L\over 4\pi} \int_{0}^{\infty} dy \ y\ln Q(iy)  \; .
\label{xxx}
\end{equation}
As we have already mentioned, this expression is valid
for conductors of arbitrary shape, as long as there is
translational invariance along the $z$-axis.  The role
of the point-matching method will be to provide an explicit expression
for the function $Q$.

\begin{figure}[ht]
\centering
\includegraphics[width=4.cm]{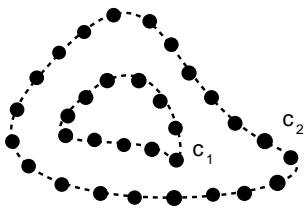}
\caption{Sections of the waveguides, indicating the points where we impose boundary conditions on
each conductor}
\label{fig2}
\end{figure}

A general solution
of the scalar Helmholtz equation (\ref{helmho})
can be written as:
\begin{equation}
R(r,\theta) =  \sum_{m=-\infty}^{\infty} [A_m J_m(\lambda r) +
B_m H^{(1)}_m(\lambda r)] e^{i m \theta},
\end{equation}
where $(r,\theta)$ are the polar coordinates, and $J_m$
and $H^{(1)}_m$ are the $m$-th order Bessel functions. The constants $A_m$ and $B_m$
are  determined by the boundary conditions.
For TM modes, the function  $R$ must verify
Dirichlet boundary conditions on each conductor  ${C}_1$
and ${C}_2$ (for TE modes, one should impose Neumann boundary conditions).
The key point is to impose the boundary conditions
on a finite number of points,  as 
it is esquematically shown in Fig.\ref{fig2}. Therefore, for TM modes we have
\begin{eqnarray}
0=\sum_{m=-S}^{S} [A_m J_m(\lambda r_p) +
B_m H^{(1)}_m(\lambda r_p)] e^{i m \theta_p},
\label{c1} \\
0=\sum_{m=-S}^{S} [A_m J_m(\lambda r_q) +
B_m H^{(1)}_m(\lambda r_q)] e^{i m \theta_q},\label{c2}
\end{eqnarray}
where $(r_p,\theta_p)$ are points of the curve ${C}_1$ and $(r_q,\theta_q)$
belong to the curve ${C}_2$.
We assume that with a finite
number of terms, say $2S+1$, we can acquire a desired
computational accuracy in the solution. Thus, we should
satify Eq.(\ref{c1}) and  Eq.(\ref{c2}) at $2S+1$ points  on
${C}_1$ and  ${C}_2$.

After imposing boundary conditions, we can write, in matrix form, the
set of linear equations for the coefficients $A_m$ and $B_m$ in Eqs. (\ref{c1}) and (\ref{c2})
\begin{eqnarray}
 M_1 A + M_2 B &=& 0 ,\nonumber \\
 N_1 A + N_2 B &=& 0, \label{setmatriz}
\end{eqnarray}
where $(M_1)_{pm}= J_m(\lambda r_p) e^{i m \theta_p}$,
$(M_2)_{pm}= H^{(1)}_m(\lambda r_p) e^{i m \theta_p}$,
$(N_1)_{qm}= J_m(\lambda r_q) e^{i m \theta_q}$ and
$(N_2)_{qm}= H^{(1)}_m(\lambda r_q) e^{i m \theta_q}$.
For this system to have non trivial solutions, we shall ask that the
determinant be zero. Thus, assuming that all matrices are square (i.e there are
$2S+1$ points on each curve),
\begin{equation}
 det\bigg[ \begin{array}{c c}
 M_1 & M_2  \\
N_1 & N_2
      \end{array}
\bigg] = det N_2 \cdot det (M_1 - M_2 N_2^{-1} N_1)=0 ~.
\end{equation}
This equation determines the eigenfrequencies associated to the geometry,
and the usual approach in classical electromagnetism would be to 
find numerically its roots.  However, as already mentioned, in order to compute
the Casimir energy it is not necessary to find 
each eigenvalue. It is far more efficient to use the argument theorem,
and obtain the sum over all eigenvalues as an integral in the complex plane.
 As this is in general a  divergent quantity, we will compute the 
 interaction zero-point energy by  substracting 
the energy of the same geometric configuration but with a very large outer conductor.
This means that we will have another set of equations similar
to Eqs.(\ref{setmatriz}) but evaluated on a different outer surface ${C}_2^\infty$
(in which $r_q$ is replaced by $ \alpha r_q$, with $\alpha\gg 1$)
\begin{equation}
 det\bigg[ \begin{array}{c c}
 M_1 & M_2  \\
N_1^{\infty} & N_2^{\infty}
      \end{array}
\bigg] \approx det\bigg[ \begin{array}{c c}
 M_1 & M_2  \\
N_1^{\infty} & 0
      \end{array}
\bigg]= det [M_2 N_1^{\infty}]=0 ~.
\end{equation}
In the equation above we have taken into account that, when using the argument
theorem, all matrices will be evaluated on the imaginary axis. In this case,
the different matrices become proportional to the modified Bessel functions
\begin{equation}
J_m(i y r_p)=i^mI_m(y r_p); \,\,\,\, H_m^{(1)}(i y r_p)=\frac{2}{\pi}(-i)^mK_m(y r_p)\, .
\label{modified}
\end{equation}
After these substitutions,
it is easy to check that the matrix $N_2^\infty$
vanishes in the limit $\alpha\rightarrow\infty$.
Then, in order to compute the finite interaction energy
with the argument theorem,
the relevant determinant is
\begin{equation}
 det  [(M_1 - M_2 N_2^{-1} N_1)M_2^{-1} (N_1^{\infty})^{-1}].
\end{equation}
Neglecting the self-energies of each
configuration (that do not contribute to the interaction
energy between the conductors), we obtain the expression
for the function  $Q$ to be included in the
expression for the Casimir interaction
energy in Eq. (\ref{xxx}):
\begin{eqnarray}
Q(i y) &=& det [I- M_1 N_1^{-1} N_2  M_2^{-1}]\nonumber\\
&=& det [I-\tilde M_1 \tilde N_1^{-1}\tilde N_2 \tilde M_2^{-1}],
\label{Qpmm}
\end{eqnarray}
where
\begin{eqnarray}
(\tilde M_1)_{pm}&=& I_m(y r_p) e^{i m \theta_p}\nonumber\\
(\tilde M_2)_{pm}&=& K_m(y r_p) e^{i m \theta_p}\nonumber\\
(\tilde N_1)_{qm}&=& I_m(y r_q) e^{i m \theta_q}\nonumber\\
(\tilde N_2)_{qm}&=& K_m(y r_q) e^{i m \theta_q}\,\, .
\label{matrixfin}
\end{eqnarray}
Note that the factors that multiply the modified Bessel functions
in Eq.(\ref{modified}) cancel out in the function $Q$.

There are  similar expressions for the TE modes, in which each Bessel function
is replaced by its derivative. The full Casimir interaction energy is given
by the sum of the TE and TM contributions. 

As a first example, we consider the case of having a configuration
of concentric cylinders. In that case, the points on curve ${C}_1$ are all
on a circle of radii $r_p=a$, i.e. we are choosing points
of coordinate ($a, \theta_p$). Similarly, points on the
outer cylinder of radii $b$ will be ($b, \theta_q$)
since we can choose points with the same
polar angle $\theta_p$ but of different radial distance.
In such a case, the matrices can be factorized as
\begin{eqnarray}
(\tilde M_1)_{pm}&=& \Theta_{pm} I_m(y a),~(\tilde M_2)_{pm}=\Theta_{pm} K_m(y a) ,\nonumber \\
(\tilde N_1)_{pm}&=& \Theta_{pm} I_m(y b),~(\tilde N_2)_{pm}=\Theta_{pm} K_m(y b) \,\, .
\label{matrixcc}
\end{eqnarray}
Here $\Theta$ is the matrix that contains the angular contribution
\begin{equation}
\Theta_{pm} = \exp\{i m \theta_p\},
\label{thetacc}
\end{equation}
where, in the case of circular sections, we have $\theta_p = 2\pi p/(2 S + 1)$.  
Inserting Eqs.(\ref{matrixcc}) and (\ref{thetacc}) into Eq.(\ref{Qpmm})
we find
\begin{equation}
Q(iy) = det\bigg[1 - \frac{I(y a)K(y b)}{I(y b) K(y a)}\bigg],
\end{equation}
where we introduced diagonal matrices $I$ and $K$.
In this way, we reobtain the known expression for the TM Casimir energy
 for concentric cylinders \cite{Mazzitelli2003}:
\begin{equation}
 E_{12}^{cc,TM} =  \frac{L}{4\pi} \int_0^{\infty} dy~y~\ln\bigg(
\prod_m \bigg[ 1 - \frac{I_m(y a)K_m(y b)}{I_m(y b) K_m(y a)} \bigg]
\bigg) \nonumber
\end{equation}
For the TE modes we the Casimir energy is given by  a similar expression,  in which 
each Bessel function is replaced by its derivative.

We have developed a numerical Fortran routine
in order to evaluate the Casimir interaction energy for a two-conductor
waveguide of arbitrary cross sections. With a two dimensional
mesh and the use of $2S+1$ points on each curve,
we define the cross section
of the waveguide. Once the points are selected, we impose
Dirichlet and Neumann boundary conditions on each of them
and define the ${\tilde M}$'s and ${\tilde N}$'s matrices. Then
we use a standard Fortran routine to invert the matrices and
define the matrix whose determinant gives the Q function (see Eq.(\ref{Qpmm}))
to be included in Eq.(\ref{xxx}).
Finally, we calculate the determinant
and  perform an integration
over all values of $y$. The parameters used by the program are:
the number of points used to define the geometry $2S+1$, the integration 
limit ($y_{\rm max}$)
and the precision desired. 

In principle, the energy can be computed accurately if
enough points are taken. This quantity rules the bigger order
of the Bessel functions involved in our simulations, and this
means that, sometimes, special precautions have to be
taken to maintain accuracy of computation 
(such as the approximation
for small argument for the modified Bessel functions).
However, we should remark that, though its limitation in the
number of points selected, this technique is of great use
for evaluating the TE and TM contributions to the energy.
In the case of smooth geometries, we shall see that the number
of points selected is enough to reproduce results in excellent
agreement with previous calculations.

Finally, we would like to mention that the choice of the
functions used to described the general solution to the Helmholtz
equation (Bessel and complex exponential functions in our case)
depends on the cross section of the waveguide. Other geometries
could  also be worked out using a different set of complete functions.

\section{Testing the method: Eccentric Cylinders}
\label{ecc}

We shall begin by applying the point-matching approach
to the case of two eccentric cylinders. In this case,
we can check our simulations against our previous numerical
work \cite{PRD}. Therein, we numerically evaluated the Casimir
interaction energy for two eccentric cylinders using the formula

\begin{eqnarray}
E_{12} &= & \frac{L}{4 \pi} \int_0^{\infty} dy
~y \bigg[{\rm ln}(M^{\rm TE}(y))
+ {\rm ln}(M^{\rm TM}(y))\bigg]  \nonumber
\end{eqnarray}
where $M^{\rm TM}(y)={\rm det}[\delta_{n p}-
A_{n,p}^{\rm TM}]$ and
$M^{\rm TE}(y)={\rm det}[\delta_{n p}-
A_{n,p}^{\rm TE}]$. The matrices $A_{n,p}^{\rm TM}$
and $A_{n,p}^{\rm TE}$ were analytically derived
in a previous work \cite{NJP}.
Herein, we shall
evaluate the Casimir interaction energy by the
use of the point-matching
technique. We will follow the notation of Ref.\cite{NJP}, denoting 
by $a$ and $b$ the radii of the inner and outer cylinders, respectively,
and by $\epsilon$ the eccentricity of the configuration (distance
between the axes of the cylinders). The adimensional 
quantitites $\delta =\epsilon/a$ and $\alpha = b/a$ will be useful
to describe the numerical results. 

\begin{figure}[h!t]
\centering
\includegraphics[width=6.5cm]{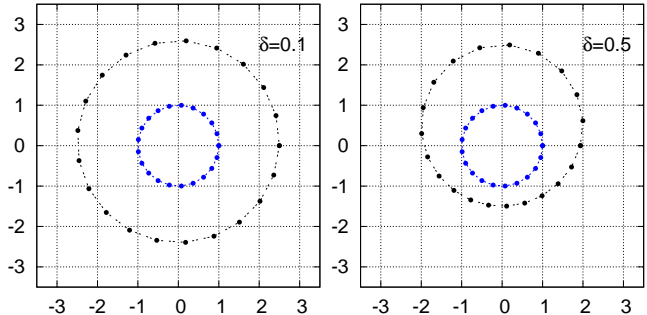}
\caption{Numerical mesh and points used in the
point matching method for simulating the boundaries
between two eccentric cylinders. In this case,
we show two different values of $\alpha$ and 
of the eccentricity $\delta$. For simplicity, for each point on the 
inner cylinder we choose a point on the outer cylinder with the 
same angular coordinate. }
\label{fig3}
\end{figure}

\begin{figure}[h!t]
\centering
\includegraphics[width=8.6cm]{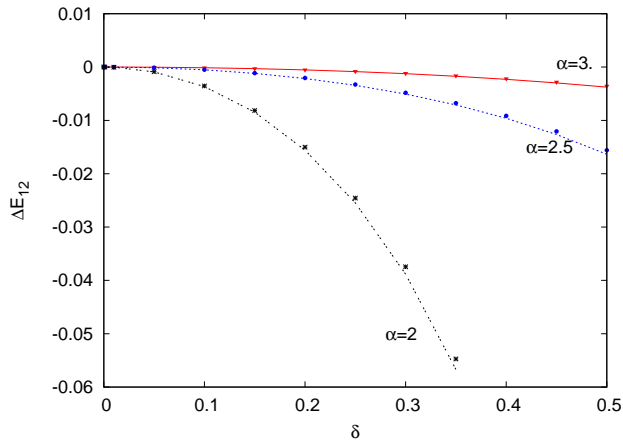}
\caption{Comparison between the point-matching method (set in the plot as
new) and the numerical method used in Ref.\cite{PRD} (set as old)
for numerically evaluating the Casimir energy as a
function of the eccentricity $\delta$ for different values of
$\alpha=b/a$. $\Delta E_{12}$ refers to the energy difference
between the eccentric and the concentric configurations. Energies
are measured in units of $L/4\pi a^2$.} \label{fig4}
\end{figure}

\begin{figure}[h!t]
\centering
\includegraphics[width=8.6cm]{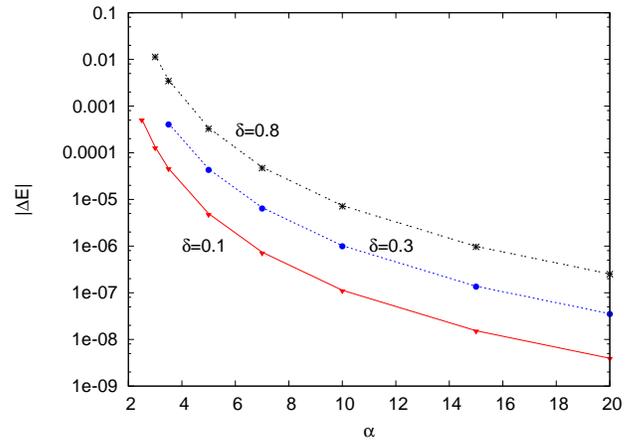}
\caption{Comparison between the new and old approaches for
the evaluation of the Casimir energy differences as a function of
$\alpha=b/a$ for different values of the eccentricity $\delta$.
Energies are measured in units of $L/4\pi a^2$.} \label{fig5}
\end{figure}

\begin{figure}[h!t]
\centering
\includegraphics[width=8.6cm]{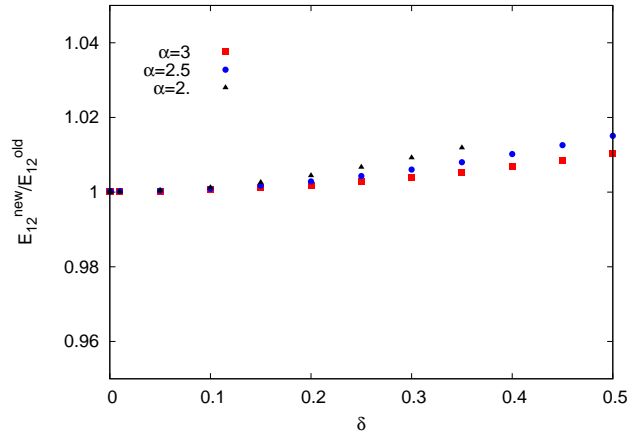}
\caption{Ratio between the Casimir energies evaluated using the
new and the old approaches, as a function of the eccentricity
$\delta$ for different values of $\alpha=b/a$.} 
\label{fig6}
\end{figure}

\begin{figure}[h!t]
\centering
\includegraphics[width=8.6cm]{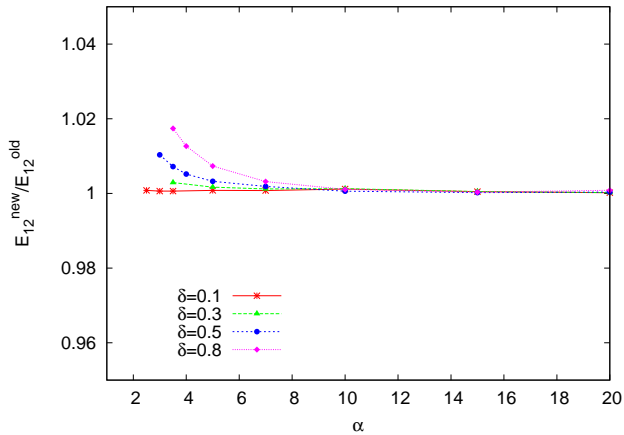}
\caption{Ratio between the Casimir energies evaluated using the
new and the old approaches, as a function of $\alpha = b/a$ for
three different values of $\delta$. 
Energies are
measured in units of $L/4\pi a^2$.} \label{fig7}
\end{figure}

\begin{figure}[h!t]
\centering
\includegraphics[width=8.6cm]{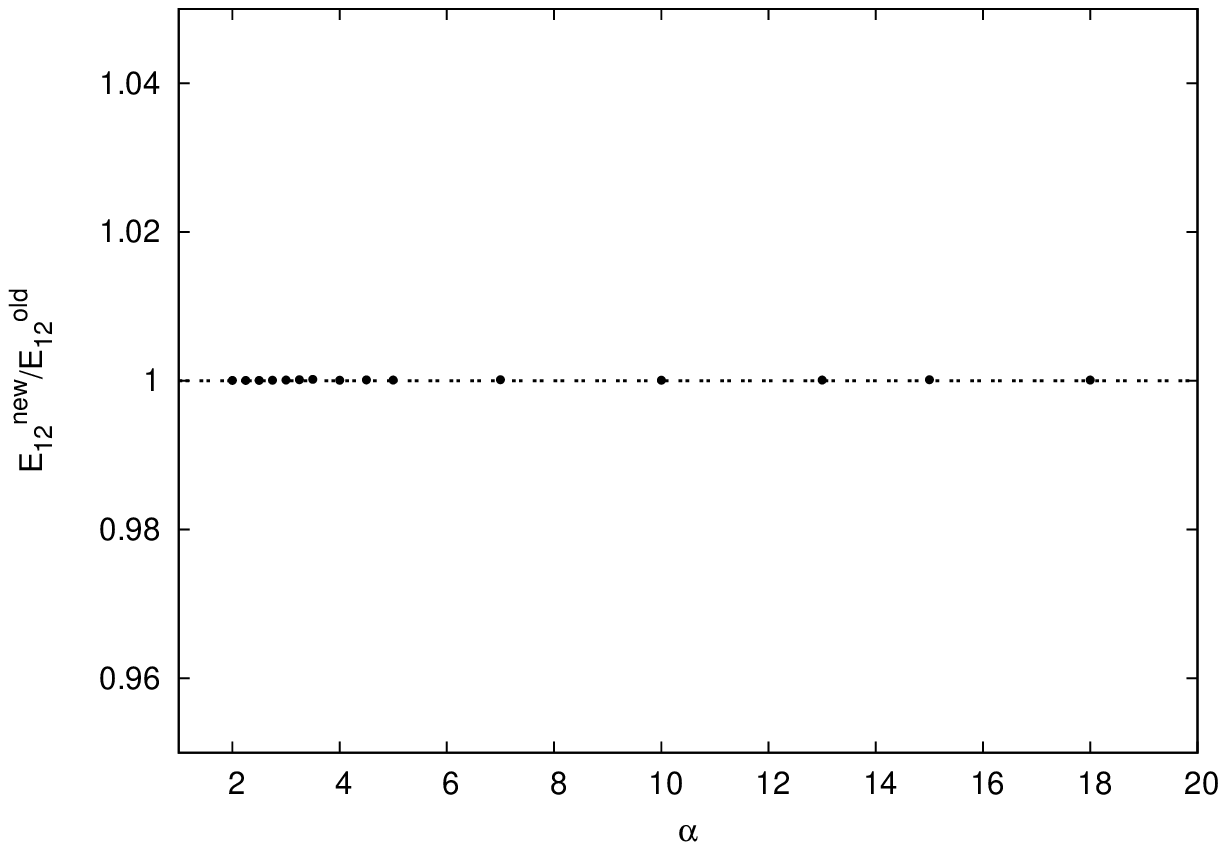}
\caption{Comparison between the new and old approach for numerically
evaluating the Casimir interaction energy in the concentric case
($\delta = 0$), as a function of $\alpha=b/a$. The point-matching
approach is greatly accurate in this case.} \label{fig8}
\end{figure}

The comparison between the new and old approaches for the evaluation
of the Casimir interaction energy between eccentric cylinders was
done for several runs with the same parameters in both programs. In
this case we used a mesh of 21 points for simulating each cylinder, and
matrices of 21x21 in our old approach.
In this way, we are able to reobtain Figs. 2 and 3 of our previous
work \cite{PRD} as can be seen in Figs. \ref{fig4} and \ref{fig5},
where we plot the interaction energy for eccentric cylinders for
different values of the radii and the eccentricity, using both
methods. It is easy to note in Figs. \ref{fig4}, \ref{fig5}, and
 \ref{fig6} that the difference among
them is less than the $2\%$ in the case of $\alpha \geq 2$ and $\delta \leq 0.5$. The
error can be blamed to the bigger quantity of algebraic operations
contained in the new approach. It is worth mentioning that adding
one more point to the point-matching method, i.e. having 23 points in
each circle, represents a variation in our result of only
$0.000004\%$ for $\alpha=3$ and $\delta=0.1$ and $0.0004\%$ for
$\alpha=2$ and $\delta=0.1$. Thus, we see the great agreement
between both approaches. As can be seen from Figs.\ref{fig6} 
and \ref{fig7}, small deviations appear for large eccentricities 
and small $\alpha$. We expect the precision in these cases 
to be improved by considering grids with a larger number 
and/or a non-uniform distribution of points on the surfaces.

In Fig. \ref{fig8} we show the comparison between the new and old
approaches for  concentric ($\delta = 0$) cylinders, as a function of $\alpha$. In
this case, the agreement between the full numerical and the
point-matching approaches is even better.

\section{Cylindrical rack and pinion}
\label{rack}

\begin{figure}[h!t]
\centering
\texttt{}\includegraphics[width=6.5cm]{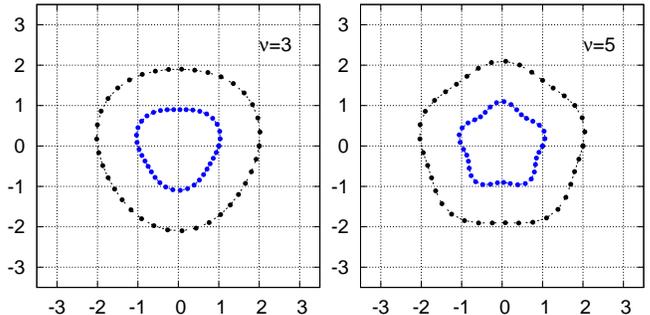}
\caption{Numerical mesh and points used in the point matching method
for simulating the boundaries between two concentric cylinders with
harmonic corrugations. In this case, we show $\alpha=2,\,\,\phi_0=0$,  for two
different values of $\nu$. } \label{fig9}
\end{figure}

When two concentric cylinders have corrugations, the vacuum energy
produces a torque that could,
in principle, make one cylinder rotate with respect to the other.
This ``cylindrical rack and pinion'' has been  proposed in Ref. \cite{brasilia}, where the torque
has been computed using the proximity force approximation. It was further analyzed
in \cite{CaveroI},  where the authors obtained
perturbative results for Dirichlet boundary conditions in the
limit of small amplitude corrugations.
In this Section, we numerically evaluate the Casimir interaction
energy for two concentric corrugated cylinders. The cylinders have
radii $a$ and $b$, and we will denote by $r_-=b-a$ the mean distance between them and 
by $r_+=a+b$ the sum of the radii. 
As in the previous Section, we will use the notation $\alpha = b/a$.
In Fig. \ref{fig9} we show two
geometries with different frequencies associated to the
corrugations: on the left side, $\nu=3$ and on the right side
$\nu=5$, both for $\alpha=2$. The points in the mesh are described by the following
functions:
\begin{eqnarray}
h_a(\theta) &=& h \sin(\nu \theta) \nonumber \\
h_b(\theta) &=& h \sin(\nu \theta + \phi_0),
\end{eqnarray}
where $h$ is the corrugation amplitude and $\nu$ is the
frequency associated with these corrugations. The Casimir torque
can be calculated by taking the derivative of the interaction
energy with respect to the shifted angle ${\cal T}= - \partial E_{12}/
\partial \phi_0$.
In the case of the TM mode (Dirichlet boundary conditions), details of the
perturbative calculation can be found in Ref.\cite{CaveroI}. Therein, the
authors
obtained an analytical expression for the Casimir interaction
energy as a function of the
angle $\phi_0$ for small $h$,  which reads 
\begin{equation}
\frac{E_{12}}{\pi r_+ L}= \cos(\nu \phi_0) \frac{\pi^2}{240 r_-^5}
h^2 B_\nu^{(2)D}(y),
\label{milton}
\end{equation}
where $y=r_-/r_+$ and $B_\nu^{(2)D}(y)$ is given by 
\begin{eqnarray}
B_\nu^{(2)D}(y) &=& \frac{15}{\pi^4} \sum_{m=-\infty}^{+\infty} 8y^3\int_0^\infty x dx \nonumber \\
& \times & \frac{4y^2}{(1-y^2)}\frac{1}{D_m(y,x)}\frac{1}{D_{m + \nu}(y,x)}.
\end{eqnarray}
The functions $D_m$ are given by
\begin{eqnarray}
 D_m(y ; x) &=& I_m(x[1 + y])  K_m(x[1 - y])  \nonumber \\
&-& I_m(x[1 - y])  K_m(x[1 + y]).
\end{eqnarray}

\begin{figure}[h!t]
\centering
\includegraphics[width=8.6cm]{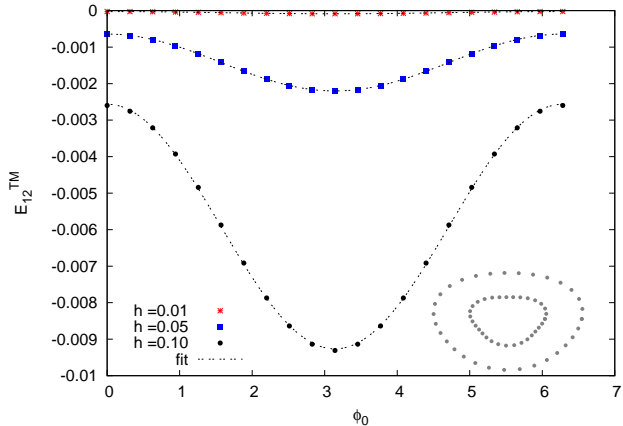}
\caption{Casimir interaction energy (TM modes)
as a function of
$\phi_0$ for $\alpha =2$ and different
values of perturbation $\tilde h=h/a$. 
The different shaped dots are the numerical data obtained
with our programe while the line represents the numerical fit of each curve.
Energies are measured in units of $L/a^2$.}
\label{fig10}
\end{figure}

In Fig.\ref{fig10} we show the numerical evaluation of the
TM Casimir interaction energy for this geometry. 
The plot shows the results obtained using our
point matching approach with  $\alpha=2$ and
corrugation frequency $\nu=3$, for different values of the amplitude
of the corrugation $h$.  As expected the amplitude of the oscillations grows with $h$. 
For each value of $h$ we have performed a numerical fit of the data in order
to compare with the analytical prediction presented in Eq.(\ref{milton}). With dotted lines 
we have plotted the fit
$y(x)=A*cos(x)$ for each curve in Fig.\ref{fig10}. The agreement
between dots and dotted lines is extremely good.

In Table \ref{tabla}, we can see the comparison between the
analytical prediction and our numerical results for different
values of $\alpha$, $\nu$ and $\tilde h = h/a$.
For each simulation we have used a mesh of 37-41
points,  depending the value of $\tilde h$. For $\alpha=2$ and
$\tilde h=0.1$, having $31$ or $37$ points selected to define
each boundary has a relative error of $0.00004\%$ in the final value of the
energy. For $\nu=5$, this difference is approximately $0.002\%$, as
one can anticipated (for a smoother curve less points are needed in
order to achieve the same accuracy in the result).
In the table, we can see that for small values of $\tilde h$, the values
of the amplitude of the numerical fit of the data are
extremely similar to those predicted analytically. However,
we must note that this technique does not have a constraint
on the value of $\tilde h$ so far. For bigger values of
$\tilde h$ one has to include more points in the mesh
so as to have a well defined geometry and
mantain accuracy in the results, that differ from the analyical 
predictions.

\begin{center}
\begin{table}
\begin{tabular}{c|c|c|c|c}\hline \hline
 $\nu$   &  $\alpha$  & $\tilde h$
& A (Analytical) & A (Numerical)   \\  \hline \hline
3 & 2 & 0.01 & 0.000030414 & 0.00003044 \\
3 & 2 & 0.05 & 0.000760 & 0.00077 \\
3 & 2 & 0.1 &  0.0030 & 0.0033  \\
3 & 2 & 0.3 &  0.027 & 0.078   \\ \hline \hline
3 & 3.5 & 0.01 & 0.000003925 &  0.000003928\\
3 & 3.5 & 0.05 & 0.0000981 & 0.0000993 \\
3 & 3.5 & 0.1 & 0.000392 & 0.000412\\
3 & 3.5 & 0.3 & 0.00353  & 0.00567\\ \hline \hline
5 & 2 & 0.01 & 0.00002046  & 0.00002049\\
5 & 2 & 0.05 & 0.000511 &  0.000528\\
5 & 2 & 0.1 & 0.00204 &  0.00232 \\
5 & 2 & 0.3 & 0.018 & 0.071\\
\end{tabular}
\caption{Comparison between the analytical (Eq.(\ref{milton}))
and numerical predictions (fit of the form $y(x)= A*\cos(x)$)
 of the Casimir interaction energy (Dirichlet modes)
for different configurations of the concentric corrugated cylinders.
}
\label{tabla}
\end{table}\end{center}

\begin{figure}[h!t]
\centering
\includegraphics[width=8.6cm]{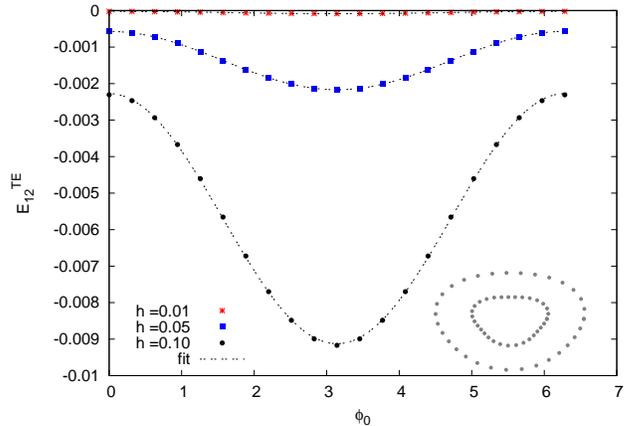}
\caption{Casimir interaction energy (TE modes)
as a function of
$\phi_0$ for a value $\alpha=2$ and  different
values of perturbation $\tilde h$.  The different shaped
dots are the numerical data obtained
with our programe while the line represents the numerical fit of each curve.
Energies are measured in units of $L/a^2$.}
\label{fig11}
\end{figure}

In addition, in Fig. \ref{fig11} we have presented the  evaluation of the
Casimir interaction energy for the Neumann (TE) modes. Therein, we see
that the behaviour is qualitatively similar to that of the Dirichlet modes, with a
different value for each numerical fit of the data points. It is
worth emphasizing that there are no analytical predictions for this
mode so far, being this the first evaluation of the TE Casimir
interaction energy for corrugated concentric
cylinders.

\begin{figure}[h!t]
\centering
\includegraphics[width=8.6cm]{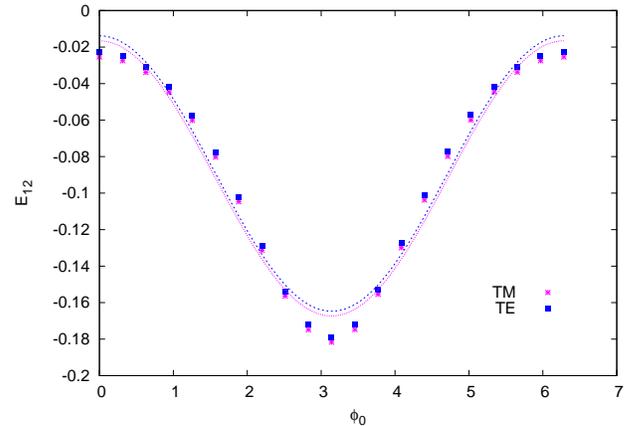}
\caption{Casimir interaction energy (TE and TM modes) as a function
of $\phi_0$ for $\alpha=2$, $\nu = 3$ and $\tilde h = 0.3$. The
different shaped dots are the numerical data obtained by our
programme while the line represents the numerical fit of each curve.
In this case, the plot shows that the exact result cannot be fitted
by a function $y(x) = A * \cos(x)$. Energies are measured in units
of $L/a^2$.} \label{fig12}
\end{figure}

\begin{figure}[h!t]
\centering
\includegraphics[width=8.6cm]{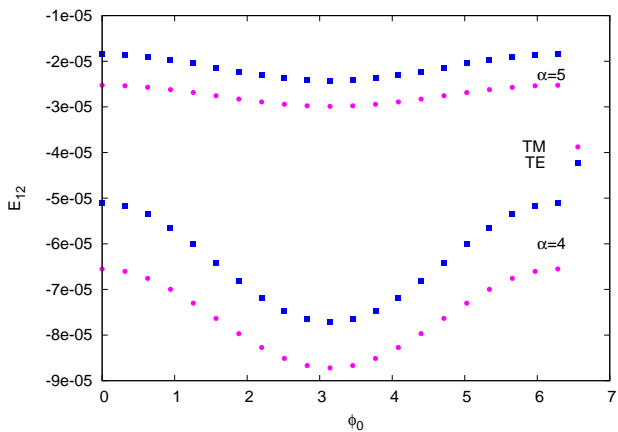}
\caption{TM and TE contributions to the Casimir interaction energy
in the concentric corrugated case. Dirichlet (TM) contribution is
bigger than the Neumann (TE) one (in absolute value).} \label{fig13}
\end{figure}

It is worth to remark that, when the amplitude
of the corrugation is not very small, the exact results cannot be
reproduced with a simple fit of the form $y(x)= A*\cos(x)$. This is
illustrated in Fig.\ref{fig12}, where we see that, for the biggest
corrugated amplitude that we included in Table \ref{tabla} ($\tilde h = 0.3$),
the exact result differs from the cosine function.

Finally, in Fig.\ref{fig13} we show the different contributions of the TM and TE
modes to the interaction energy for bigger values of $\alpha$.
As expected from previous results \cite{NJP}, for large values of $\alpha$, 
Dirichlet contribution is bigger (in absolute value) than the 
Neumann one.

\section{Outer conductors with focal lines}

Some time ago, there was a conjeture \cite{ford} based on a geometric optics 
approximation, about the possibility of focusing vacuum fluctuations in 
parabolic mirrors. It was argued that a parabolic mirror is capable of 
focusing the vacuum modes of the quantized electromagnetic field,  therefore
creating large physical effects near the mirror's focus. The physical 
manifestation of this focusing is a growth in the energy density and mean-squared 
electric field as the focus is approached. In particular, the energy density 
would diverge as the inverse fourth power of the distance from the focus in the 
case of perfect conductivity. These results would imply that the focused vacuum 
fluctuations will enhance Casimir forces on atoms or other particles near the 
focus. The sign of the force could draw particles into the vecinity of the 
focus \cite{ford}. 

With this motivation, in this Section we shall evaluate the Casimir interaction  
energy for configurations in which the outer conducting shell has a cross
section that contains focal points. 

\subsection{Cylinder inside an ellipse}
\label{elipses}

To begin with, we will 
compute the Casimir interaction energy between one small inner cylinder and an 
outer ellipse,
by the use of the point matching method. We will denote by $a$ the radius of the inner cylinder, by $b_1$ and $b_2$ 
the minor and major semiaxes of the ellipse, respectively, and by $f$ the distance between
the foci and the center of the ellipse. The coordinates of the center of the cylinder with respect to the center of the ellipse
will be $(\epsilon_x,\epsilon_y)$. We will use an additional tilde to denote adimensional quantities, i.e distances in units
of $a$: $\tilde b_i=b_i/a\, , \tilde f=f/a$, etc.

\begin{figure}[h!t]
\centering
\includegraphics[width=8.6cm]{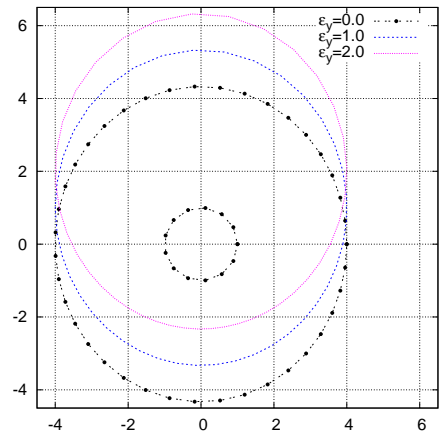}
\caption{Mesh used to evaluate
the boundaries condition in the framework of the point-matching
method. We represent a small centered cylinder and
 vary the position of the outer ellipse. Parameters
used: minor ellipse semiaxis $\tilde b_1=4$,
major ellipse semiaxis $\tilde b_2=4.33$ and focal position $\tilde f=1.66$.}
\label{fig14}
\end{figure}

\begin{figure}[h!t]
\centering
\includegraphics[width=8.6cm]{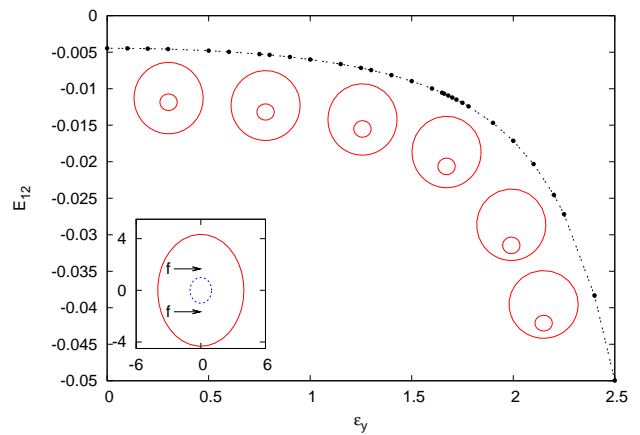}
\caption{Numerical evaluation of the Casimir interaction energy for
an inner cylinder an eccentric outer ellipse, as a function
of the position of the cylinder along the vertical axis. Energies are measured in units of $L/a^2$.}
\label{fig15}
\end{figure}

\begin{figure}[h!t]
\centering
\includegraphics[width=8.6cm]{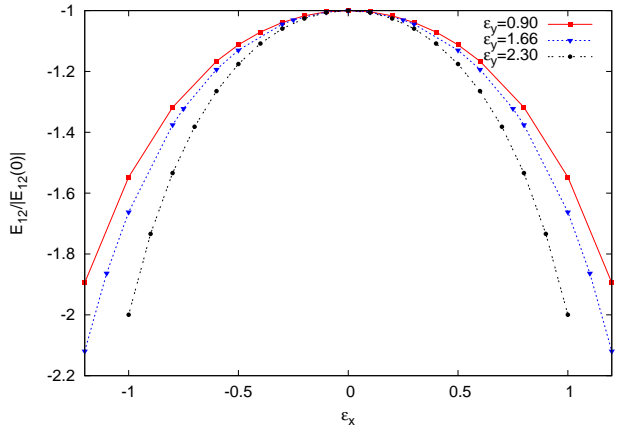}
\caption{Moving the inner cylinder along the minor semiaxis
the ellipse we show that the center of the ellipse is an unstable 
equilibrium position.
This is shown for different positions of the inner cylinder along
the major semiaxis. Energies are normalized to the value of $\vert E_{12}(\epsilon_x =0)\vert$. } \label{fig16}
\end{figure}

For this configuration, we use a mesh like the one presented in Fig.
\ref{fig14}, where we show an inner cylinder, 
and an outer ellipse with semiaxes $\tilde b_1 = 4$ and $\tilde b_2 = 4.33$. The ellipse
has two focal points at  $\tilde f= 1.66$. We present the
results for the Casimir energy in Figs.\ref{fig15} and \ref{fig16}.

From Fig.\ref{fig15} it is possible to see that there is an 
unstable equilibrium position at the origin under displacements of the 
inner cylinder along the (vertical)  $\epsilon_y$ direction. As expected, 
it is also possible to check that the energy grows as well as 
the cylinder gets closer to the surface of the outer ellipse. Fig.\ref{fig15} also 
shows a monotonic behaviour of the energy as a function of the position, even when
passing through the focus.  
Otherwise, Fig.\ref{fig16} shows the unstable equilibrium position at the origin when 
moving the inner cylinder in the (horizontal) $\epsilon_x$ direction. It is 
also important to stress that, when considering horizontal displacements
at a fixed vertical position, the higher the altitude of the inner cylinder, 
the narrow the inverted potential in Fig.\ref{fig16}.

\subsection{Cylinder inside a Parabola}
\label{parabola}

Following the same idea than in the previous subsection, here we describe the numerical 
computation of  the Casimir interaction
energy for a small inner cylinder with circular cross section, 
inside a large cylinder with parabolic section.
\begin{figure}[h!t]
\centering
\includegraphics[width=8.6cm]{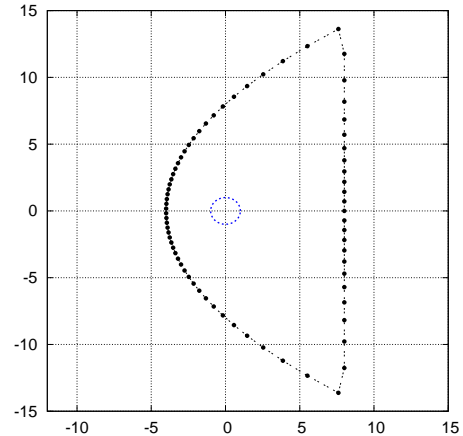}
\caption{Mesh used to evaluate
the boundaries condition in the framework of the point-matching
method. We represent a small centered cylinder and
 vary the value of the position of the outer parabola. Parameter
used  $\tilde f=4$.}
\label{fig17}
\end{figure}
In Fig.\ref{fig17} we show the two dimensional cross 
section of the mesh used to simulate
this geometry and impose the boundary conditions.
The parameters
for this case are the radius of the inner cylinder, $a$, and 
the focal distance of the parabola, $f$. As before we introduce $\tilde f=f/a$.  

\begin{figure}[h!t]
\centering
\includegraphics[width=8.6cm]{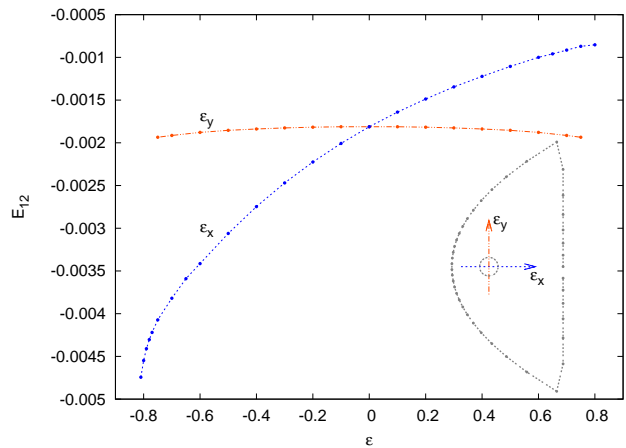}
\caption{Numerical evaluation of the Casimir interaction energy
between an inner cylinder inside a bigger parabola. We move the inner
cylinder both in directions $\epsilon_x$ and  $\epsilon_y$. Parameter
used: $\tilde f=4$. Energies are measured in units of $L/a^2$.}
\label{fig18}
\end{figure}

In Fig.\ref{fig18},
we show the behaviour of the Casimir
interaction energy with the position of the inner cylinder.
The number of points used is approximately 40.
It is easy to see that the
bigger increasement of the energy (and therefore of the force)
is in the direction labeled as $\epsilon_x$.
We can also note an increasement in
the energy as the inner cylinder gets closer to the vertex of
the parabola (in the $\epsilon_x$ direction), a biproduct of the proximity
between the inner cylinder surface and the one corresponding to the
parabola. On the other hand, as the behaviour of the energy is
symmetric in the $\epsilon_y$ direction,
the force $F_y$ vanishes on the horizontal axis. There is also an unstable equilibrium 
position at a particular point on this axis, as it is suggested by Fig.\ref{fig18}.
As in the previous example, the energy and the force do not have a special
behaviour near the focus.

\section{Final remarks}

In this paper we presented a new numerical method to compute the vacuum energy for
arbitrary geometries with translational invariance. The method is based on the use of the point-matching approach,
in which the boundary conditions are imposed on a discrete set of points on the surfaces of
the conducting bodies. This approach is combined with the argument theorem, in order to
trade the sum over eigenvalues for an integral in the complex plane.

After testing our method against previous results, we have computed the Casimir interaction energy for new geometries. In the case of the cylindrical rack and pinion described in Section IV, we have seen
that, for corrugations of small amplitude, the numerical results for the energy show a harmonic dependence with the shifted angle, and a quadratic dependence with the amplitude of corrugations, in agreement with previous analytic perturbative evaluations.
This behaviour disappears for larger amplitudes, where the exact results show more pronounced peaks.

In Section V we computed the Casimir interaction energy between an inner cylinder and
outer surfaces with focal lines. The motivation for looking at these configurations was to see whether
there was a non trivial behaviour of the energy near the focus or not. For both cases considered
(ellipses and parabolas), we have found that the energy and forces are monotonic across the focal lines.
This may be a peculiarity of the geometries with translational invariance considered here.
We have also confirmed  the existence  of unstable equilibrium positions of the inner cylinder, that coincide with  the location suggested by simple geometric arguments.  

The examples discussed in this paper  illustrate the simplicity and power  of this approach.
We have used a straightforward  version of the point-matching method, with a naive choice 
of the points on the curves  (note that in all examples we have chosen pair of  points with the
same angular coordinate with respect to the inner cylinder). For less symmetric configurations,
and when the surfaces of both conductors are closer to each other, it will be necessary to
consider grids with a larger number of points, and to  optimize their positions. 
As in the applications to acoustic or classical electromagnetism, special care must be taken for surfaces with pronounced edges, clefts or "handles", where the point-matching technique may not be accurate to determine the eigenfrequencies.

 It is  certainly possible
to go beyond the geometries considered here. Still considering geometries with translation invariance along the $z$-axis, it would be possible to analyze waveguides with more than two conducting bodies.
The generalization for non-perfect mirrors is also possible. For the case of scalar fields  satisfying  arbitrary matching conditions on the surfaces (rather than Dirichlet or Neumann boundary conditions), the generalization is relatively
straightforward. However, for the electromagnetic field the
calculation is more cumbersome, since in general it will be not possible to treat independently the TE and TM modes.

The point-matching approach can also be generalized to three dimensional compact objects. The starting point in this case 
would be the solutions of the three dimensional Helmholtz equation, written in terms of an adequate basis
(products of spherical Bessel functions and spherical harmonics). We expect these geometries to require much more computational effort, and a more sophisticated method to optimize the choice of the grid of points on which the boundary conditions are imposed. 

\label{conc}
\acknowledgments
This work has been supported by CONICET, UBA and ANPCyT,
Argentina. P.I.V. would like to thank the hospitality of the Barcelona Supercomputing Centre,
where part of this work was done.

\end{document}